\documentclass[reprint, amsmath,amssymb, aps]{revtex4-2}
\usepackage{graphicx}
\usepackage{xcolor}
\usepackage{float}%we can use the more powerful H, which will ensure the figure is kept in place
\usepackage{dcolumn}% Align table columns on decimal point
\usepackage{bm}% bold math
\usepackage{natbib}
\usepackage{hyperref}% add hypertext capabilities

\begin{document}

\preprint{APS/123-QED}

\title{Causal description of marginally trapped surfaces in D-dimensions}

\author{Konka Raviteja}
\email{konka.raviteja@gmail.com}
\author{Asrarul Haque}
\email{ahaque@hyderabad.bits-pilani.ac.in}
\author{Sashideep Gutti}
\email{sashideep@hyderabad.bits-pilani.ac.in}

\affiliation{Department of Physics, Birla Institute of Technology and Sciences-Pilani \\Hyderabad, 500078, India}

\date{\today} 

\begin{abstract}
In this paper, we analyze the causal aspects of evolving marginally trapped surfaces in a D-dimensional spherically symmetric spacetime, sourced by perfect fluid with a cosmological constant. The norm of the normal to the marginally trapped tube is shown to be the product of lie derivatives of the expansion parameter of future outgoing null rays along the incoming and outgoing null directions. We obtain a closed form expression for this norm in terms of principal density, pressure, areal radius and cosmological constant. For the case of a homogeneous fluid distribution, we obtain a simple formula for determining the causal nature of the evolving horizons. We obtain the causal phase portraits and highlight the critical radius. We identify many solutions where the causal signature of the marginally trapped tube or marginally anti-trapped tube is always null despite having an evolving area. These solutions don't comply with the standard inner and outer horizon classification for degenerate horizons. We propose an alternate prescription for this classification of these degenerate horizons.

\end{abstract}

\maketitle

%\tableofcontents
\section{Introduction} \label{sec:intro}

The study of blackholes has engaged a great many physicists over the years. The event horizon of stationary blackholes has been of considerable interest to theoretical physicists owing to it's connections with blackhole thermodynamics, where the notion of temperature and entropy are well defined. The notion of these thermodynamic variables is still in the nascent stage for non-stationary evolving horizons. Due to the detection of gravitational waves and their promises in the field of observational astronomy, the evolving horizons and their understanding is of crucial importance to theoretical physicists and observational astronomers. 
\\

Ashtekar and Badrikrishnan \cite{ashtekar2002dynamical,ashtekar2003dynamical} have constructed the dynamical horizon as a three dimensional spacelike hypersurface and the timelike membrane as three dimensional timelike hypersurface which are foliated by marginally trapped surfaces (MTS). They derive area laws for the dynamical horizons where the area increases monotonically for a choice of time, the area decreases monotonically for timelike membranes. There are therefore separate area laws for dynamical horizons and timelike membranes. These timelike membranes and dynamical horizons can be seen in the Lemaitre-Tolman-Bondi collapse models, where we observe a transition from timelike membrane to a dynamical horizon as one moves from centre of the star to the outside and all these horizons are inside the event horizon. In general it is possible for an evolving horizons to transition from spacelike to timelike or vice versa. Some of these transitions can be seen in the models constructed in these papers. \cite{booth2005marginally,helou2017causal,raviteja2020aspects,sherif2019some,chatterjee2020marginally}
\\

Stitching together the area laws of dynamical horizon and timelike membrane, Bousso and Engelhardt \cite{bousso2015new,bousso2015proof} have proposed a new area law for (future or past) holographic screens \cite{bousso1999holography}, where for certain allowed transitions of holographic screens the area increases or decreases monotonically throughout.
\\

The lack of sufficient number of analytically tractable models is one of the difficulties in the study of evolving horizons. In general the complexity of the Einstein equations makes the analysis accessible mostly via numerical study, however in simpler settings there are a few models that are accessible analytically. In our previous work \cite{raviteja2020aspects} we have constructed one such model with evolving dust in a D-dimensions and found timelike to spacelike transitions and vice versa. We saw for the evolving horizons in this model that even with the transitions, the area evolution is monotonic (either increases or decreases) along these horizons, as argued in \cite{bousso2015new}.
\\

The plan of the present paper is to generalize the analysis compared to our previous work \cite{raviteja2020aspects} by considering perfect fluid with pressure as the matter source. The dust model we studied considerably simplified the metric and the Einstein equations and hence it is was possible to extract the solutions of Einstein equations. For a perfect fluid evolution in D-dimensions, constructing analytical solutions of Einstein equations is not possible. However the causal nature of evolving horizons is still analytically obtainable just from the structure of the Einstein equations which is shown in section II. These formulae are expressed in terms of the principal densities, pressures, cosmological constant and Area radius. In section III we look at the homogeneous case with an equation of state for the perfect fluid. We obtain an important geometric result for the causal nature of the evolving horizon in the homogeneous setting. If cosmological constant is not zero, the spacetime is partitioned into two regions based on the Area radius, separated by a critical area radius. We also show the causal phase portraits for the evolving horizons illustrating the critical radius at which a causal transition occurs. We also find solutions of evolving marginally trapped horizons that do not comply with the horizon classification given by Hayward \cite{hayward1994general}. We resolve these by introducing tie-breaker vector fields. This analysis qualitatively gives us all transitions in evolving horizons in D-dimensional spherically symmetric spacetimes sourced by (homogeneous) perfect fluids, with and without cosmological constant.

\section{D-dimensional spherically symmetric inhomogeneous Perfect fluid evolution}

The metric for a $(D = n+2)$ dimensional spherically symmetric spacetime is of the form
    \begin{equation}\label{metric}
    ds^2 = - e^{\sigma(t,r)}dt^2 + e^{\lambda(t,r)}dr^2 + R^2 (t,r)~d{\Omega}_{n}^{2}
    \end{equation}
where $d{\Omega}_{n}^{2}$ is the metric on unit $n$ dimensional sphere with $(\theta_1, \theta_2, ..., \theta_n)$ as the angular coordinates, $t$ is the time coordinate, r is the comoving radial coordinate and $R(t,r)$ is the areal radius (we will also refer to this as `physical radius')  of the n-dimensional sphere. The matter we are considering here is a perfect fluid whose energy-momentum tensor is
    \begin{equation}
    T_{\mu\nu} = (\rho(t,r) + p(t,r))u_{\mu}u_{\nu} + p(t,r) g_{\mu\nu}   
    \end{equation}
and the four velocity in comoving coordinates is
    \begin{equation}
    u^{\mu} = (e^{-\frac{\sigma}{2}},0,0,...,0)    
    \end{equation}
with $u^{\mu} u_{\mu} = -1$. The non-zero components of the energy-momentum tensor are listed below
    \begin{eqnarray*}
    &&
    T_{00} = \rho e^{\sigma}
    \\ &&
    T_{11} = p e^{\lambda} 
    \\ &&
    T_{22} = p R^2
    \\ &&
    T_{(l+1 ~ l+1)} = sin^2{\theta_{(l-1)}} T_{(ll)} 
    \end{eqnarray*}
where $l$ takes values from 2 to n. From the Bianchi identities 
    \begin{equation}
    {T^{\mu\nu}}_{;\nu} = 0
    \end{equation}
we get the following relations
    \begin{equation} \label{timebianchi}
    \dot{\rho} + \frac{(\rho + p)}{2} \bigg(\frac{2n\dot{R}}{R} + \dot{\lambda} \bigg) = 0
    \end{equation}
    \begin{equation} \label{radialbianchi}
    p' + \frac{\sigma'}{2} (p + \rho) = 0
    \end{equation}
where the dot represents a derivative with time coordinate and prime represents derivative with the comoving radial coordinate.
\\

For a nonzero cosmological constant ($\Lambda \neq 0$), the Einstein equations are
    \begin{equation}
    G_{\mu\nu} + \Lambda g_{\mu\nu} = \kappa T_{\mu\nu}
    \end{equation}
here $\kappa$ is a constant and is related to gravitational constant $G_n$, ($\kappa$ = 8$\pi G_n$). With these conditions we evaluate the left hand side components of the Einstein equation $(G_{\mu\nu} + \Lambda g_{\mu\nu})$ which are summarized below
    \begin{eqnarray}
    && G_{00}+\Lambda g_{00} = \frac{e^{-\lambda}}{R^2} \bigg[ \frac{n(n-1)}{2} (e^{\lambda + \sigma}+ e^{\lambda} \dot{R}^2 - e^{\sigma} R'^{2})  \nonumber \\ && + \frac{n}{2} R (-2 R'' e^{\sigma} + e^{\sigma} R' \lambda' + e^{\lambda}  \dot{R} \dot{\lambda}) - \Lambda e^{\lambda + \sigma} R^{2} \bigg]
    \end{eqnarray}
    \begin{equation} \label{G01}
    G_{01} +\Lambda g_{01}= \frac{n}{2}\frac{(R' \dot{\lambda} - 2 \dot{R}' +  \sigma'\dot{R} )}{R} 
    \end{equation}
    \begin{eqnarray}
    && G_{11}+\Lambda g_{11} = \frac{e^{-\sigma}}{R^{2}} \bigg[ - \frac{n(n-1)}{2} (e^{\lambda + \sigma}+ e^{\lambda} \dot{R}^2 - e^{\sigma} R'^{2})  \nonumber \\ && + \frac{n}{2} R ( e^{\sigma} R' \sigma' + e^{\lambda}(\dot{R}\dot{\sigma} - {\color{black} 2\ddot{R}})) + \Lambda e^{\lambda + \sigma} R^{2} \bigg] 
    \end{eqnarray}
    \begin{eqnarray}
    && G_{22}+\Lambda g_{22} =  \frac{e^{-(\lambda+\sigma)}}{4}  \bigg[ 2(n-1)(n-2) ( e^{\sigma} R'^{2} - e^{\lambda +\sigma} - e^{\lambda}\dot{R}^2) \nonumber
    \\ &&
    - 2(n-1)R (e^{\sigma} R'(\lambda' - \sigma') - 2  e^{\sigma} R''+ e^{\lambda}(\dot{R}(\dot{\lambda} - \dot{\sigma}) + 2\ddot{R})) \nonumber 
    \\ && 
    + R^{2}(4e^{\lambda + \sigma} \Lambda - e^{\lambda}(2\ddot{\lambda} +\dot{\lambda}^2 - \dot{\lambda}\dot{\sigma}) +  e^{\lambda}(2\sigma''+ {\sigma'}^2 - \lambda' \sigma'))  \bigg] \nonumber
    \end{eqnarray}
The other nonzero relations are given by 
    \begin{equation}
    G_{(j+1~j+1)} = sin^2{\theta_{(j-1)}} G_{(jj)}
    \end{equation}
where j takes values from 2 to n. So from the $G_{01} = 0$ Einstein equation we get
    \begin{equation} \label{simpleG01}
    R'\dot{\lambda} - 2 \dot{R}' + \sigma'\dot{R} = 0
    \end{equation}
from the $G_{00} = \kappa \rho e^{\sigma}$ equation we have
    \begin{eqnarray} \label{G00}
    && 
    \frac{n(n-1)}{2} (e^{\lambda + \sigma}+ e^{\lambda} \dot{R}^2 - e^{\sigma} R'^{2})  
    \\ && 
    + \frac{n}{2} R (-2 R'' e^{\sigma} + e^{\sigma} R' \lambda' + e^{\lambda}  \dot{R} \dot{\lambda}) =  (\kappa \rho +\Lambda) R^2 e^{\lambda + \sigma} \nonumber
    \end{eqnarray}
and $G_{11} = \kappa p e^{\lambda}$ equation we get
    \begin{eqnarray} \label{G11}
    && 
    -\frac{n(n-1)}{2} (e^{\lambda + \sigma}+ e^{\lambda} \dot{R}^2 - e^{\sigma} R'^{2})
    \\ && 
    + \frac{n}{2} R ( e^{\sigma} R' \sigma' + e^{\lambda}(\dot{R}\dot{\sigma} -  2\ddot{R}) ) =  (\kappa p - \Lambda) R^2 e^{\lambda + \sigma} \nonumber
    \end{eqnarray}
The following two expressions will be useful in the subsequent calculations done in the paper. The sum (\ref{G00}) + (\ref{G11}) gives,
    \begin{eqnarray} \label{G00+G11}
    && \frac{n}{2} \bigg( e^{\sigma}R'(\sigma' + \lambda') + e^{\lambda}\dot{R}(\dot{\sigma}+\dot{\lambda}) - 2 (e^{\lambda}\ddot{R} + e^{\sigma} R'')   \bigg)      \nonumber
    \\ 
    && = \kappa (\rho + p) e^{(\lambda + \sigma)} R
    \end{eqnarray}
similarly the difference (\ref{G00}) - (\ref{G11}) gives us
    \begin{eqnarray} \label{G00-G11}
    && 
    n(n-1) (e^{\lambda + \sigma}+ e^{\lambda} \dot{R}^2 - e^{\sigma} R'^{2})  
    \\ && 
    + \frac{nR}{2}  ( e^{\sigma}(R' \lambda' - 2 R'' - R' \sigma') - e^{\lambda}(\dot{R}\dot{\sigma} -  2\ddot{R} - \dot{R} \dot{\lambda} )) \nonumber
    \\ &&
    =  (\kappa(\rho - p) + 2\Lambda) R^2 e^{\lambda + \sigma} \nonumber
    \end{eqnarray}

\subsection{Causal Nature}

For the metric (\ref{metric}) the future outgoing radial null vector is
    \begin{equation}
    k^a = (e^{-\frac{\sigma}{2}},e^{-\frac{\lambda}{2}},0,0....,0)
    \end{equation}
and the future incoming incoming radial null vector is
    \begin{equation}
    l^a = (e^{-\frac{\sigma}{2}},-e^{-\frac{\lambda}{2}},0,0....,0)
    \end{equation}
these null vectors are normalized as 
    \begin{equation*}
    g_{ab}k^{a}l^{b} = -2
    \end{equation*}
with these null vectors the induced metric on a codimension 2 hypersurface is given by
    \begin{equation*}
    h_{ab} = g_{ab} + \frac{1}{2}(k_{q}l_{b} + l_{a}k_{b})
    \end{equation*}
so the expansion for the congruence of outgoing null rays is 
    \begin{equation}
    \Theta_{k} =  h^{ab} \nabla_{a} k_{b} = \frac{n}{R} \bigg( e^{-(\frac{\sigma}{2})} \dot{R} + e^{-(\frac{\lambda}{2})} R' \bigg) 
    \end{equation}
and the expansion for the congruence of incoming null rays is 
    \begin{equation}
    \Theta_{l} =  h^{ab} \nabla_{a} k_{b} = \frac{n}{R} \bigg( e^{-(\frac{\sigma}{2})} \dot{R} - e^{-(\frac{\lambda}{2})} R' \bigg) 
    \end{equation}
 We restate the standard definition of a marginally trapped surface (MTS) to be a codimension 2 submanifold $\Sigma$ whose expansion of null congruence $\Theta_k$ generated by the outgoing radial null vector $k^a$ vanishes everywhere ($\Theta_k$ = 0) on $\Sigma$ and $\Theta_l$ which is expansion of null congruence generated by incoming radial null vector $l_a$ is completely negative on $\Sigma$ ($\Theta_l < 0$). A curve that is foliated by the MTS are called as marginally trapped tube (MTT).
\\

Similarly we define marginally antitrapped surfaces (MATS) as codimension 2 submanifold $\Xi$ whose expansion of incoming radial null congruence $\Theta_l$ vanishes everywhere ($\Theta_l$ = 0) on $\Xi$ and the expansion of outgoing radial null congruence $\Theta_k$ is completely positive on $\Xi$ ($\Theta_k > 0$). A curve that is foliated by the MATS are called as marginally antitrapped tube (MATT).
\\

Note that for the MTS and MATS, we have the inequality relations $\Theta_{l} < 0$ and $\Theta_{k} > 0$ respectively and in this model both these inequalities give the same condition, that is 
    \begin{equation}
    \frac{R'}{e^{\frac{\lambda}{2}}} > 0
    \end{equation}
This condition holds true if there is no shell crossing and it ensures that the comoving coordinate system does not become singular except at the curvature singularity. We assume that this condition holds at both the MTS and MATS. 
\\

The MTT and MATT are codimension 1 submanifolds and the causal nature of these tubes could be completely timelike (timelike membranes), completely spacelike (dynamical horizons), completely null (as will be demonstrated in the subsequent sections) or have a mixed signature where causal transitions are allowed. There are a couple of methods that can be used for evaluating the causal nature of MTT and MATT. One method utilizes the norm of the normal to the curves that are foliated by MTS and MATS, this was outlined in \cite{raviteja2020aspects} and is suited better for spherically symmetric case. We first evaluate the norm of the normal to the surface $\Theta_{k}=c$. Since the surface is trivial in the angular coordinates, we focus on the $t$,$r$ sector. This norm can be expressed as a product of Lie derivatives of $\Theta_{k}$ or $\Theta_{l}$ along the incoming and outgoing null rays $l^{a}$, $k^{a}$ on MTT or MATT.
\\

The surface $\Theta_{k} = c$ will be a curve foliated by MTS when $c = 0$, so the curve is a MTT. The norm of this curve is given by 
    \begin{equation*}
    \begin{split}
    \beta_{k} = e^{-\lambda} n_{r}^{2} - e^{-\sigma} n_{t}^{2} = (e^{\frac{-\lambda}{2}}n_{r} + e^{\frac{-\sigma}{2}}n_{t}) (e^{\frac{-\lambda}{2}}n_{r} - e^{\frac{-\sigma}{2}}n_{t})
    \end{split}
    \end{equation*}
where $n_{t}$ is the time derivative of $\Theta_{k}$ and $n_{r}$ is the radial derivative of $\Theta_{k}$ (we assume metric functions are smooth and the derivatives exist). The norm for the theta curve when evaluated at $\Theta_{k} = 0$ is
    \begin{equation}
    \begin{split}
    \beta_{k} = \bigg(\bigg(\frac{\partial_{r} \Theta_{k}}{e^{\frac{\lambda}{2}}}  + \frac{\partial_{t} \Theta_{k}}{e^{\frac{\sigma}{2}}}  \bigg) \bigg( \frac{\partial_{r} \Theta_{k}}{e^{\frac{\lambda}{2}}}  - \frac{\partial_{t} \Theta_{k}}{e^{\frac{\sigma}{2}}} \bigg) \bigg) \bigg|_{\Theta_{k}=0}
    \end{split}
    \end{equation}
We show that the above expression can be written as a product of two Lie derivatives of $\Theta_{k}$. The lie derivative of $\Theta_{k}$ with respect to the outgoing null vector $k^{a}$ is
    \begin{equation} \label{explkok}
    \pounds_{k} \Theta_{k} = k^{a} \nabla_{a} \Theta_{k} = e^{\frac{-\sigma}{2}} \partial_{t} \Theta_{k} + e^{\frac{-\lambda}{2}} \partial_{r} \Theta_{k}
    \end{equation}
This lie derivative has to be evaluated at $\Theta_{k} = 0$. From the relation (\ref{simpleG01}) the lie derivative becomes 
    \begin{equation*}
    \begin{split}
    \pounds_{k} \Theta_{k} \bigg|_{\Theta_{k} = 0} & =  \frac{n e^{-\sigma}}{R} (\ddot{R} - \frac{\dot{\sigma}\dot{R}}{2}) + \frac{n e^{-\lambda}}{R} (R'' - \frac{\lambda' R'}{2})
    \\
    & + \frac{n}{2R}  (\sigma' \dot{R} + R'\dot{\lambda}) e^{-\frac{({\lambda+\sigma})}{2}}  
    \end{split}
    \end{equation*}
which is reduced further by using the relation (\ref{G00+G11}) that are evaluated after substituting $\Theta_k=0$, we get
    \begin{equation}
    \pounds_{k} \Theta_{k} \bigg|_{\Theta_{k} = 0} = - \kappa (\rho + p)
    \end{equation}
The other lie derivative that is to be evaluated at $\Theta_{k} = 0$ is 
    \begin{equation} \label{expllok}
    \pounds_{l} \Theta_{k} = l^{a} \nabla_{a} \Theta_{k} =  e^{\frac{-\sigma}{2}} \partial_{t} \Theta_{k} - e^{\frac{-\lambda}{2}} \partial_{r} \Theta_{k}
    \end{equation}
again using the relation (\ref{simpleG01}) the lie derivative (evaluated at $\Theta_k=0$) becomes
    \begin{equation*}
    \begin{split}
    \pounds_{l} \Theta_{k} \bigg|_{\Theta_{k} = 0} & = \frac{n e^{-\sigma}}{R} (\ddot{R} - \frac{\dot{\sigma}\dot{R}}{2}) - \frac{n e^{-\lambda}}{R} (R'' - \frac{\lambda' R'}{2})
    \\
    & + \frac{n}{2R}  (\sigma' \dot{R} - R'\dot{\lambda}) e^{-\frac{({\lambda+\sigma})}{2}}  
    \end{split}
    \end{equation*}
now using the relation (\ref{G00-G11}) we get
    \begin{equation}
    \pounds_{l} \Theta_{k} \bigg|_{\Theta_{k} = 0} = \kappa (\rho - p) + 2\Lambda - \frac{n (n-1)}{R^2}
    \end{equation}
The other set of lie derivatives for the expansions of the incoming null rays which are to be evaluated when $\Theta_{l} = 0$ are
    \begin{equation} \label{explkol}
    \pounds_{k} \Theta_{l} = k^{a} \nabla_{a} \Theta_{l} = e^{\frac{-\sigma}{2}} \partial_{t} \Theta_{l} + e^{\frac{-\lambda}{2}} \partial_{r} \Theta_{l}
    \end{equation}
using the relation (\ref{simpleG01})
    \begin{equation*}
    \begin{split}
    \pounds_{k} \Theta_{l} \bigg|_{\Theta_{l} = 0} = \frac{n}{R} \bigg( &  e^{-\sigma} (\ddot{R} - \frac{\dot{\sigma}\dot{R}}{2}) - e^{-\lambda} (R'' - \frac{\lambda' R'}{2}) \bigg)
    \\
    & - \frac{n}{2R}  (\sigma' \dot{R} - R'\dot{\lambda}) e^{-\frac{({\lambda+\sigma})}{2}} 
    \end{split}
    \end{equation*}
using the relation (\ref{G00-G11}) we get
    \begin{equation}
    \pounds_{k} \Theta_{l} \bigg|_{\Theta_{l} = 0} = \kappa (\rho - p) + 2\Lambda - \frac{n (n-1)}{R^2}
    \end{equation}
The next lie derivative is
    \begin{equation} \label{expllol}
    \pounds_{l} \Theta_{l} = l^{a} \nabla_{a} \Theta_{l} = e^{\frac{-\sigma}{2}} \partial_{t} \Theta_{l} - e^{\frac{-\lambda}{2}} \partial_{r} \Theta_{l}
    \end{equation}
using the relation (\ref{simpleG01})
    \begin{equation*}
    \begin{split}
    \pounds_{l} \Theta_{l} \bigg|_{\Theta_{l} = 0} = \frac{n}{R} \bigg( &  e^{-\sigma} (\ddot{R} - \frac{\dot{\sigma}\dot{R}}{2}) + e^{-\lambda} (R'' - \frac{\lambda' R'}{2}) \bigg)
    \\
    & - \frac{n}{2R}  (\sigma' \dot{R} + R'\dot{\lambda}) e^{-\frac{({\lambda+\sigma})}{2}} 
    \end{split}
    \end{equation*}
using the relation (\ref{G00+G11}), we get
    \begin{equation}
    \pounds_{l} \Theta_{l} \bigg|_{\Theta_{l} = 0} = - \kappa (\rho + p).
    \end{equation}
We can therefore express the norm of the normal to a MTT (\ref{betak}) in terms of product of Lie derivatives as
    \begin{equation}
    \beta_{k} = - \bigg( (\pounds_{k} \Theta_{k}) (\pounds_{l} \Theta_{k}) \bigg) \bigg|_{\Theta_{k} = 0} 
    \label{productoflie}
    \end{equation}
So the norm  for the MTT is 
    \begin{equation} \label{betak}
    \beta_{k} = \kappa (\rho + p) \bigg( \kappa (\rho - p) + 2\Lambda - \frac{n (n-1)}{R^2} \bigg)
    \end{equation}
Similarly if we consider the curve $\Theta_{l}=0$ which is foliated by MATS and represents a MATT, it's norm of the normal to the curve can be expressed as the product of Lie derivatives as
    \begin{equation*}
    \beta_{l} = - \bigg( (\pounds_{l} \Theta_{l}) (\pounds_{k} \Theta_{l}) \bigg) \bigg|_{\Theta_{l} = 0}
    \end{equation*}
which is 
    \begin{equation} \label{betal}
    \beta_{l} = \kappa (\rho + p) \bigg( \kappa (\rho - p) + 2\Lambda - \frac{n (n-1)}{R^2} \bigg)
    \end{equation}
The norms of the normal to MTT and MATT therefore turn out to be proportional to product of lie derivatives and is given by the same expression.
\\

The causal nature of MTT and MATT are also found using the ratio of lie derivatives which represents the causal nature of the tangent to the curves that are foliated by MTS or MATS, the proof of which is shown in \cite{hayward1994general,dreyer2003introduction}. The ratio of the lie derivatives evaluated at $\Theta_{k} = 0$ determine the causal nature of a MTT, which is  
    \begin{equation} \label{alphak}
    \alpha_{k} = \frac{\pounds_{k} \Theta_{k}}{\pounds_{l} \Theta_{k}} \bigg|_{\Theta_{k} = 0} = \frac{- \kappa (\rho + p)}{\kappa (\rho - p) + 2\Lambda - \frac{n (n-1)}{R^2}}
    \end{equation}
and ratio of the lie derivatives evaluated at $\Theta_{l} = 0$ determines the causal nature of MATT, which is 
    \begin{equation} \label{alphal}
    \alpha_{l} = \frac{\pounds_{l} \Theta_{l}}{\pounds_{k} \Theta_{l}} \bigg|_{\Theta_{l} = 0} = \frac{- \kappa (\rho + p)}{\kappa (\rho - p) + 2\Lambda - \frac{n (n-1)}{R^2}}
    \end{equation}
Notice that the causal nature of both MTT and MATT are again determined by the same expression.
\\

Also we can see that $\beta \equiv (\beta_{k},\beta_{l})$ and $\alpha \equiv (\alpha_{k},\alpha_{l})$ are opposite in sign, which is to be expected since $\beta$ indicates the causal nature of the normal of the curves of MTS and MATS while $\alpha$ indicates the causal nature of the tangent to the curves of MTS and MATS.
\\

If we assume that the matter obeys the null energy condition, then the sign of $\beta$ or $\alpha$ is completely decided by the expression,
    \begin{equation} \label{conditioninhomogeneous}
    \pounds_{l} \Theta_{k} = \kappa (\rho - p) + 2\Lambda - \frac{n (n-1)}{R^2} = \pounds_{k} \Theta_{l}
    \end{equation}
    
Note that we were able to obtain closed form expressions for lie derivative of MTS and MATS just by using the structure of Einstein equations. This is always possible with spherical symmetric metric which can be seen from the focusing equations given in \cite{hayward1994general}. These focusing equations are shown to be the part of a system of equations which describe the dual-null dynamics of the Einstein field equations \cite{hayward1993dual}. One can perform a similar construction for D-dimensions, but in the current context, it is simpler to derive the results using the Einstein equations. 

\section{Homogeneous Fluid distribution}

We now look at MTS and MATS in homogeneous fluids. These models find their applications in cosmology and also in stellar collapse scenarios. The advantage of these models is that the equations become greatly simplified and sometimes the solutions are analytically tractable. For the metric (\ref{metric}) with homogeneous fluid distribution where both pressure and matter density depend only on time that is $p = p(t)$ and $\rho = \rho(t)$. We also assume an equation of state relation between pressure and density
    \begin{equation} \label{eos}
    p = \omega \rho.
    \end{equation}
A minimally coupled scalar field $\phi$ in a general FRW spacetime depends only on the comoving time $\phi = \phi(t)$ and can be treated as a perfect fluid (\cite{madsen1985note,madsen1988scalar,faraoni2012correspondence,semiz2012comment}). Its pressure and energy density are
    \begin{equation}
    p(t) = \frac{1}{2}\dot{\phi}^2 -  V(\phi) 
    \end{equation}
    \begin{equation}
    \rho(t) = \frac{1}{2}\dot{\phi}^2 -  V(\phi) 
    \end{equation}
For minimally coupled scalar fields the null energy condition is satisfied. The equation of state parameter is given by
    \begin{equation}
    \omega = \frac{\frac{1}{2}\dot{\phi}^2 -  V(\phi)}{\frac{1}{2}\dot{\phi}^2 + V(\phi)}
    \end{equation}
where $\dot{\phi}$ is a time derivative of $\phi$ and $V(\phi)$ is the potential energy. For free scalar field that is $V(\phi) = 0$, we have $\omega = 1$ and vanishing kinetic energy $\dot{\phi} = 0$, we have $\omega = -1$. Hence we are interested in the range of $\omega = [-1,1]$ as quite a number of equations of state of interest can modeled by the minimally coupled scalar field.
\\

We can obtain interesting closed form expressions to describe the causal nature of the MTT and MATT using the above equation of state. The Bianchi identities for the homogeneous energy-momentum tensor are
    \begin{equation} \label{homtimebi}
    \dot{\rho} + \frac{(\rho + p)}{2} \bigg(\frac{2n\dot{R}}{R} + \dot{\lambda} \bigg) = 0
    \end{equation}
    \begin{equation} \label{homradialbi}
    (\rho + p) \frac{\sigma'}{2} = 0
    \end{equation}
and  for homogeneous fluid distribution in spherically symmetric metric the physical radius $R$ can be written as 
    \begin{equation} \label{sepofvariables}
    R(r,t) = a(t)~r
    \end{equation}
Using the above conditions it is well known from \cite{chatterjee1990homogeneous}  that the metric (\ref{metric}) can be brought to the standard FRW form,
    \begin{equation} \label{ndimhommetric}
    ds^2 = - dt^2 + a^{2}(t) \bigg(\frac{dr^2}{1-kr^2}  + r^2~d{\Omega}_{n}^{2} \bigg)
    \end{equation}
which is the higher dimensional spherically symmetric metric whose source is a homogeneous perfect fluid. The function $a(t)$ has the standard interpretation as the scale factor and $k$ takes the values in  $(1,0,-1)$.
\\

The Einstein equations for metric (\ref{ndimhommetric}) with homogeneous fluid distribution are,
    \begin{equation} \label{homEE1}
    \frac{n(n+1)}{2} \frac{\dot{a}^2 + k}{a^2} = \Lambda + \kappa \rho
    \end{equation}
    \begin{equation} \label{homEE2}
    \frac{n\ddot{a}}{a} + \frac{n(n-1)}{2} \frac{\dot{a}^2 + k}{a^2} = \Lambda - \kappa p 
    \end{equation}
In addition to the Einstein equations, the evolution of pressure and density can be obtained from the Bianchi identity (\ref{homtimebi}), using the relations (\ref{eos}) and (\ref{sepofvariables}), the equation can be written as,
    \begin{equation}
    \dot{\rho} + \rho \frac{\dot{a}}{a} (n + 1) (\omega + 1) = 0.
    \end{equation}
Integrating the above equation we get the mass density $\rho(t)$ as a function of the scale factor $a(t)$. 
    \begin{equation} \label{homdensity}
    \rho(t) = \frac{\rho_{o}}{a^{(n+1)(\omega+1)}}
    \end{equation}
where $\rho_{o}$ is the initial mass density.

\subsection{Causal nature of MTT and MATT in homogeneous case}

One can arrive at the causal description of the MTT or MATT for the homogeneous case directly by using the general formula for $\beta$ in (\ref{betak}, \ref{betal}) or using the expression of $\alpha$ given in (\ref{alphak}, \ref{alphal}). Instead we take the longer route and compute all the lie derivatives in the standard cosmological variables. For the metric (\ref{ndimhommetric}), the future incoming radial null vector in the given by
    \begin{equation}
    l^a = (1,-\frac{\sqrt{1 - k r^2}}{a(t)},0,0....,0)
    \end{equation}
and the future outgoing radial null vector is given by
    \begin{equation}
    k^a = (1,\frac{\sqrt{1 - k r^2}}{a(t)},0,0....,0).
    \end{equation}
These are normalized to, 
    \begin{equation*}
    g_{ab}k^{a}l^{b} = -2.
    \end{equation*}
The expansion scalar for outgoing bundle of null rays is 
    \begin{equation} \label{homthetak}
    \Theta_{k} =  h^{ab} \nabla_{a} k_{b} = \frac{n}{a(t)r} \bigg( \dot{a}(t) r + \sqrt{1 - k r^2} \bigg) ,
    \end{equation}
while the expansion scalar for incoming bundle of null rays is 
    \begin{equation} \label{homthetal}
    \Theta_{l} =  h^{ab} \nabla_{a} k_{b} = \frac{n}{a(t)r} \bigg( \dot{a}(t) r - \sqrt{1 - k r^2}  \bigg) 
    \end{equation}
we know that the norm of the normal to the $\Theta_{k} = 0$ curve can be expressed as product of Lie derivatives (\ref{productoflie}). The lie derivative of $\Theta_{k}$ with respect to the outgoing radial null vector is
    \begin{equation}
    \pounds_{k} \Theta_{k} = k^{a} \nabla_{a} \Theta_{k} = \partial_{t} \Theta_{k} + \frac{\sqrt{1 - k r^2}}{a(t)} \partial_{r} \Theta_{k}
    \end{equation}
this lie derivative has to be evaluated at $\Theta_{k}=0$, so from the expression (\ref{homthetak}) we have $\dot{a}(t) r = - (\sqrt{1 - k r^2})$, which means the above lie derivative simplifies as 
    \begin{equation*}
    \pounds_{k} \Theta_{k} \bigg|_{\Theta_{k}=0} = -\frac{nk}{a^2} -n( \frac{\dot{a}^{2}}{a^2} - \frac{\ddot{a}}{a} )
    \end{equation*}
From einstein equations (\ref{homEE1})+(\ref{homEE2}), we get the quantity 
    \begin{equation} \label{homEE1+EE2}
    n( \frac{\dot{a}^{2}}{a^2} - \frac{\ddot{a}}{a} ) = \kappa (\rho + p) -\frac{nk}{a^2}
    \end{equation}
Using this relation the lie derivative reduces to  
    \begin{equation}
    \pounds_{k} \Theta_{k} \bigg|_{\Theta_{k}=0} = -\kappa \rho ( 1 + \omega)
    \end{equation}
The other lie derivative that is to be evaluated at $\Theta_{k} = 0$ is 
    \begin{equation}
    \pounds_{l} \Theta_{k} = l^{a} \nabla_{a} \Theta_{k} =  \partial_{t} \Theta_{k} - \frac{\sqrt{1 - k r^2}}{a(t)} \partial_{r} \Theta_{k}
    \end{equation}
using the expression (\ref{homthetak}) for $\Theta_{k}=0$ gives 
    \begin{equation*}
    \pounds_{l} \Theta_{k} \bigg|_{\Theta_{k}=0} = \frac{2n}{a^2 r^2} - \frac{n k}{a^2}  -n( \frac{\dot{a}^{2}}{a^2} - \frac{\ddot{a}}{a} )
    \end{equation*}
using the relation (\ref{homEE1+EE2}) and (\ref{sepofvariables}) this is further reduced to 
    \begin{equation}
    \pounds_{l} \Theta_{k} \bigg|_{\Theta_{k}=0} = \frac{n}{2R^2}\left(3-n-\omega(n+1)+\frac{2\Lambda R^2 (1+\omega)}{n}\right)
    \end{equation}
Similarly, the lie derivatives of $\Theta_{l}$ with respect to the ingoing radial null vector evaluates to,
    \begin{equation}
    \pounds_{l} \Theta_{l} \bigg|_{\Theta_{k}=0} = -\kappa \rho ( 1 + \omega)
    \end{equation}
and
    \begin{equation}
    \pounds_{k} \Theta_{l} \bigg|_{\Theta_{k}=0} = \frac{n}{2R^2}\left(3-n-\omega(n+1)+\frac{2\Lambda R^2 (1+\omega)}{n}\right)
    \end{equation}
The norm of the normal to the curves $\Theta_{k}=0$ and $\Theta_{l}=0$ can therefore be expressed as
    \begin{equation}   \label{hombeta}
    \beta_{k} = \frac{n\kappa \rho ( 1 + \omega)}{2R^2}
    \left( 3-n-\omega(n+1)+\frac{2\Lambda R^2 (1+\omega)}{n} \right) = \beta_{l}
    \end{equation}
As described earlier the causal nature of MTT and MATT can also be found using the ratio of lie derivatives which represents the causal nature of the tangent to MTT and MATT. The ratio of the lie derivatives evaluated at $\Theta_{k} = 0$ and $\Theta_{l} = 0$ gives us
    \begin{equation} \label{homalpha}
    \alpha_{k} = \frac{-2R^2 \kappa \rho ( 1 + \omega)}{n\left(3-n-\omega(n+1)+\frac{2\Lambda R^2 (1+\omega)}{n}\right)} = \alpha_{l}
    \end{equation}
Note that we have expressed the formula in terms of physical radius $R$ instead of $a(t)r$. From the $\alpha$ and $\beta$ expressions we can see that the MTT and MATT are spacelike if $\beta<0$ ($\alpha>0$), timelike if $\beta>0$($\alpha<0$) and null if $\beta=0$ ($\alpha=\infty$).
\\

Assuming the null energy condition, the sign of the expressions for $\beta$ and $\alpha$ are completely determined by sign of the term
    \begin{equation} \label{criticalterm}
    \left( 3-n-\omega(n+1)+\frac{2\Lambda R^2 (1+\omega)}{n} \right)
    \end{equation}
as rest of the terms in the expressions are positive definite. This term in the expressions of $\beta$ and $\alpha$ does not contain any dynamical variables of the model and is expressed completely in terms of spacetime dimension ($n$), equation of the state parameter ($\omega$), cosmological constant ($\Lambda$) and physical radius ($R$) that is completely characterized by the killing vectors of the spherically symmetric spacetime. In this sense it is a geometrical result.
\\

When $\omega = 0$, which means pressure $p(t) = 0$ we recover the results form our previous work \cite{raviteja2020aspects}.

\subsection{Causal phase portraits of MTT and MATT in parameter space}

In this subsection, we look at the causal phase portrait in the parameter space ($\Lambda$, $n$, $\omega$, $R$) for MTT and MATT. Let us first look at the case for $\Lambda \neq 0$ where there exists a critical radius $R_{critical}$ at which MTT and MATT are null ($\beta = 0 = \alpha$) and it also marks a transition of these curves from spacelike region to timelike region or vice versa. 
    \begin{equation*}
    R_{critical}^{2} = \frac{n(n-3) + n\omega (n+1)}{2 \Lambda (1+\omega)}
    \end{equation*}
    \begin{figure}[H]
    \centering 
    \includegraphics[scale=0.9]{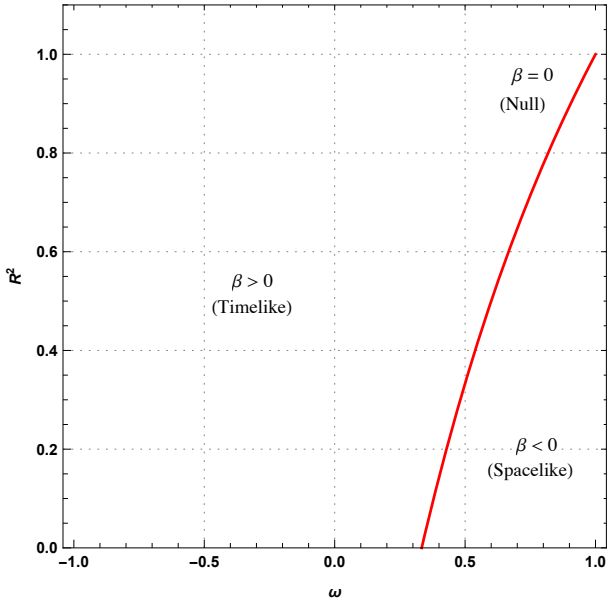}
    \caption{$R^{2}$ versus $\omega$ for MTT or MATT with ($\Lambda = 1$, D = 4)}
    \label{D4ds}
    \end{figure}
The causal phase portrait for four dimensional $(n=2)$ desitter spacetime (we choose $\Lambda =1$ for the purpose of illustration) is shown in figure (\ref{D4ds}). We have indicated regions where the MTT or MATT will be timelike or spacelike and the line indicates where they become null representing the values of $R_{critical}$ for the range of $\omega$ parameter. We see that for all the values of $\omega$ in the range [-1 to 1/3) the MTT is always timelike and is  a timelike membrane. However for any given value of $\omega$ in the range (1/3 to 1] the MTT under goes a causal transition from spacelike to timelike and the point of transition is $R_{critical}$, hence in this range we have MTT with mixed signatures. The MATTs also have the exact same behaviour.
\\

\begin{figure}[H]
    \centering 
    \includegraphics[scale=0.9]{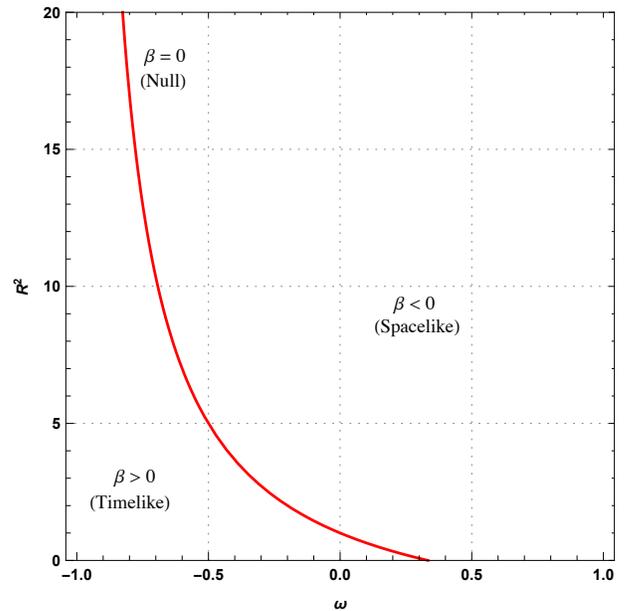}
    \caption{$R^{2}$ versus $\omega$ for MTT or MATT ($\Lambda=-1$, D = 4)}
    \label{D4ads}
    \end{figure}

The causal phase portrait for four dimensional $(n=2)$ Anti desitter spacetime $(\Lambda = -1)$ is shown in figure (\ref{D4ads}). The line again indicates where MTT and MATT become null and represents the values of $R_{critical}$ for the range of $\omega$ parameter.
For all the values of $\omega$ in the range [1 to 1/3) the MTT is always spacelike and is called a dynamical horizon. For $\omega$ in the range (1/3 to -1] the MTT undergoes a causal transition from timelike to spacelike hence the MTT have a mixed signature in this region. Again the MATT also have the exact same behaviour. Note that there are timelike membranes possible in the desitter spacetime and dynamical horizons in anti desitter spacetime, while both of these spacetime also allow for mixed signature in MTT and MATT. 
\\

    \begin{figure}[H]
    \centering 
    \includegraphics[scale=0.9]{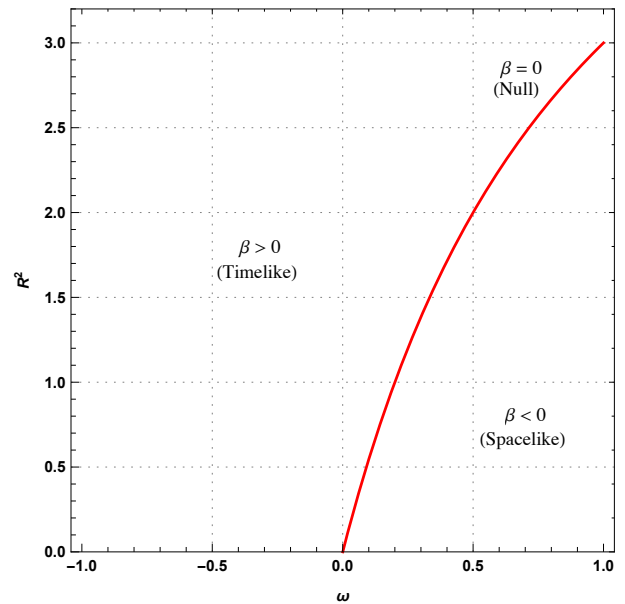}
    \caption{$R^{2}$ versus $\omega$ for MTT or MATT with ($\Lambda = 1$, D = 5)}
    \label{D5ds}
    \end{figure}
The causal phase portrait for five dimensional $(n=3)$ desitter spacetime $(\Lambda = 1)$ is shown in figure (\ref{D5ds}). Just as before, the line drawn indicates the critical radius as a function of $\omega$. The MTT is a timelike membrane for $\omega$ in the range [-1 to 0) and in the range (0 to 1] it has a mixed signature with a spacelike to timelike transition.  
\\

\begin{figure}[H]
    \centering 
    \includegraphics[scale=0.9]{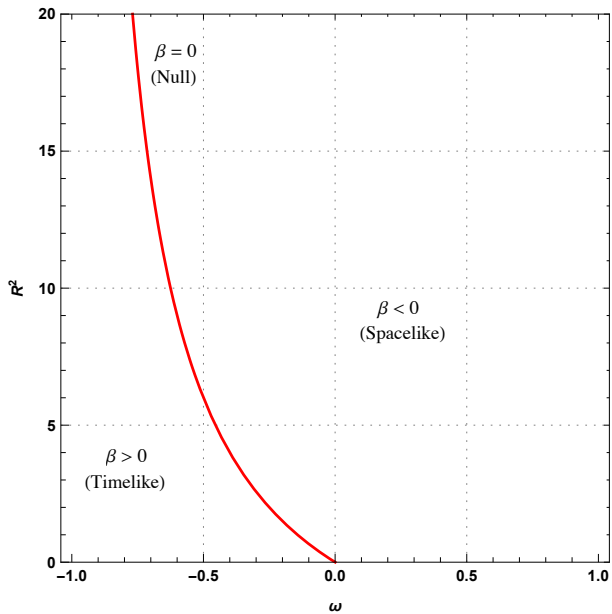}
    \caption{$R^{2}$ versus $\omega$ for MTT or MATT with ($\Lambda = -1$, D = 5)}
    \label{D5ads}
    \end{figure}

\begin{figure}[H]
    \centering 
    \includegraphics[scale=0.9]{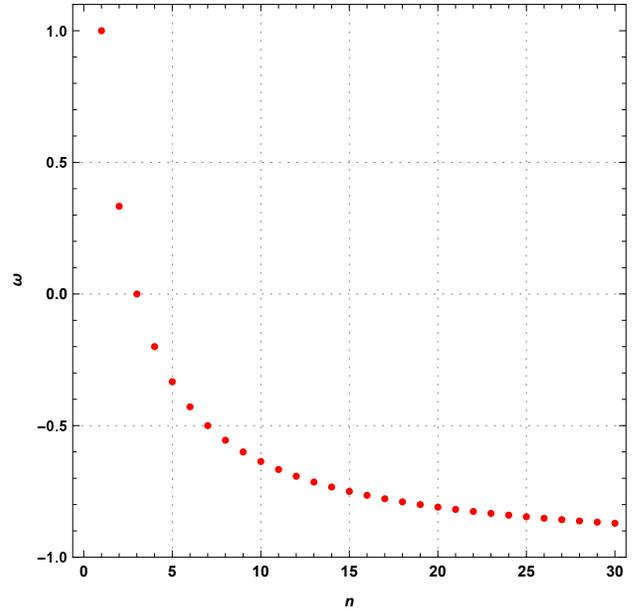}
    \caption{$\omega$ versus $n$ for MTT or MATT ($\Lambda = 0$, D = {3 to 30})}
    \label{wvsn}
    \end{figure}

The causal phase portrait for five dimensional $(n=3)$ anti-desitter spacetime $(\Lambda = -1)$ is shown in figure (\ref{D5ads}). Here again we have dynamical horizon for $\omega$ in the range [1 to 0) and a mixed signature MTT with a timelike to spacelike transition for $\omega$ values in the range (0 to -1].
Note that as the dimension$(n)$ keeps increasing the zeros of $R^2$ that is the value of $\omega$ where $R^2 = 0$ where causal transitions either start or end for MTT and MATT keeps shifting to the left with $\omega = -1$ as its asymptotic limit. This behaviour is true in all the three spacetimes (AdS, dS and flat). These values of $\omega$ are exactly the points shown in figure(\ref{wvsn}) for various dimensions ($n$).
\\

From the expression of (\ref{hombeta}) we can see that for the case of $\Lambda = 0$ that is flat spacetime, we get a critical equation of state parameter $\omega_{critical}$ where MTT and MATT are null ($\beta = 0$). Also when $\Lambda \neq 0$ we can see that the zeros of $R^2$ occur at the same value of $\omega$ as given by $\omega_{critical}$.
    \begin{equation}
    \omega_{critical} = - \frac{(n-3)}{(n+1)}
    \end{equation}
In the figure (\ref{wvsn}) we plot the critical values of the equation of state parameter in dimensions of spacetime from 3 to 30. In flat spacetime, if $\omega < (3-n)/(n+1)$ then $\beta > 0 $  so the MTT is timelike hence these are timelike memebranes. Similarly for $\omega > (3-n)/(n+1)$ we have $\beta < 0 $ that is a spacelike MTT or a dynamical horizon. At $\omega = (3-n)/(n+1)$ which is the critical equation of state parameter $(\omega_{critical})$ we have $\beta = 0 $ hence the MTT is null. Note that for $\Lambda=0$, for a given dimension $(n)$, $\beta$ is either completely timelike, completely spacelike or null depending on $\omega$, mixed causal signatures are not allowed in this case for MTT and MATT.

\subsection{Inner and outer classification of MTT and MATT}

Hayward in \cite{hayward1994general} has introduced conditions for further classifying the future and past trapping horizons as inner and outer horizons. The MTT which is foliated by MTS is equivalent to a future trapping horizon while the MATT which is foliated by MATS is equivalent to past trapping horizon. We will see that the causal nature of the MTT or MATT will already determine the inner and outer classification of horizons. So, the MTT and MATT are classified as inner and outer horizons just by their causal nature. The conditions given in \cite{hayward1994general} for inner and outer classification are
    \begin{itemize}
    \item outer horizon if $\pounds_{l}\Theta_{k} < 0$
    \item inner horizon if $\pounds_{l}\Theta_{k} > 0$
    \end{itemize}
These above conditions hold true for MTT, while for MATT the conditions are the same but instead of the lie derivative $\pounds_{l}\Theta_{k}$ we need to use the lie derivative $\pounds_{k}\Theta_{l}$, however since both these lie derivative are equal expression wise as seen in (\ref{conditioninhomogeneous}), we can just focus on the conditions stated above for the inner and outer classification for both the MTT and MATT.
\\

From the general expressions of $\beta$ and $\alpha$ under the null energy condition, we can see that $\beta > 0$ implies that $\pounds_{l} \Theta_{k} > 0$, so a timelike MTT is also an inner horizon. Similarly $\beta < 0$ implies $\pounds_{l} \Theta_{k} < 0$, so a spacelike MTT is also an outer horizon. Hence timelike MTT are equivalent to future inner trapping horizons (FITH) and spacelike MTT are equivalent to future outer trapping horizons (FOTH). Similarly timelike MATT is past inner trapping horizon (PITH) and spacelike MATT is past outer trapping horizon (POTH). Hence the causal phase portraits for MTT and MATT can now be translated to the trapping horizon classification of Hayward.  A timelike to spacelike transition is equivalent to inner horizon to outer horizon transition. Note that AdS spacetime allow only timelike to spacelike transitions or equivalently inner to outer horizons transitions, while dS spacetime allow only spacelike to timelike transitions or equivalently outer to inner horizons transitions. As illustrated in \cite{bousso2015new}, the transition from outer to inner in a monotonically evolving trapped region is better seen in a different time coordinate where the outer horizon evolves from a smaller radius to a larger radius, meeting the inner horizon that is decreasing in radius.
\\

We now consider case when $\beta=0$ everywhere. This situation occurs when $\omega = (3-n)/(n+1)$ in the absence of a cosmological constant. We see that MTT and MATT are null. If we consider $D=4$, we see that the for $\omega=1/3$, we get that $\beta=0$. This fact re-establishes the findings in \cite{ben2004penrose} where the case of $\omega=1/3$ is discussed. This equation of state is relevant for the evolution of our universe during the radiation dominated era. This implies that during the radiation dominated era (we ignore cosmological constant and other type of matter fields), the MTS or MATS is always null and its remains null for all radius $R$ during evolution with $\omega=1/3$.  An explicit example of similar evolution was shown in \cite{raviteja2020aspects} for $D=5$ and this dealt with pressure-less dust ($\omega=0$) as source of matter. In the figure (\ref{wvsn}) each dot represents the case when the MTS or MATS is completely null throughout the evolution of MTS or MATS.
\\

We see that in these cases, we have $\pounds_{l} \Theta_{k}=0$ (also for MATT $\pounds_{k} \Theta_{l}=0$),  Null MTT ($\beta=0$) are called as future degenerate horizon and null MATT ($\beta=0$) are called as past degenerate horizon where the inner and outer classification of the such null MTT or null MATT are done by going to the higher orders of lie derivative like $\pounds_{l} \pounds_{l} \Theta_{k}$, which is a prescription given by Hayward in \cite{hayward1994general}. We will see that a classification of the degenerate horizons as outer or inner for all the points of the figure (\ref{wvsn}) is non trivial, because all the higher order lie derivatives vanish at these degenerate horizons. 
\\

For illustration set $k=0$ and check the second order lie derivatives for null MTT (or MATT). For $\Lambda=0$ at $\omega_{critical}$, $\pounds_{l} \Theta_{k}=0$ for MTT and evaluating $\pounds_{l}\pounds_{l}\Theta_k$ at $\Theta_{k}=0$ gives us 
    \begin{equation}
    \pounds_{l}\pounds_{l}\Theta_k \bigg|_{\Theta_{k} = 0} = \frac{n\dddot{a}}{a}+\frac{2n\ddot{a}}{a^2r}+\frac{5n}{a^3r^3}
    \label{doublelie}
    \end{equation}
We then make use of the Einstein equations (\ref{homEE1}),(\ref{homEE2}) and the equation of state $p = \omega_{critical} ~ \rho$, which gives us
    \begin{equation}
        \ddot{a}=-\dot{a}^2/a
    \end{equation}
differentiate the above expression with time gives
    \begin{equation}
    \dddot{a}=-\frac{2\dot{a}\ddot{a}}{a}+\frac{\dot{a}^3}{a^2}
    \end{equation}
Using these relations we get to see that the equation (\ref{doublelie}) is zero that is
    \begin{equation*}
    \pounds_{l}\pounds_{l}\Theta_k \bigg|_{\Theta_{k} = 0} = 0
    \end{equation*}
Also one could check that all the higher order lie derivatives vanish for such a degenerate horizon, this is true even when $k \neq 0$. Hence the criteria of evaluating second or higher order lie derivatives for inner or outer horizon classification turns out to be inconclusive.
\\

In these kind of scenarios, we proposed a `tie-breaker' vector field in \cite{raviteja2020aspects} where we evaluate the Lie derivative along a spacelike vector field $\xi = \frac{\partial}{\partial r}$ in the comoving coordinate chart ($t,r$). The assumption for this to be valid is that there are no shell crossing singularities due to the matter evolution and therefore the vector $\xi$ points in the direction of increasing physical radius $R$. The coordinate chart is a valid coordinate system till the we reach $R=0$ which is a curvature singularity.
\\

According to this proposal for degenerate horizons, if $\pounds_{\xi} \Theta_k > 0$ then the MTS or MATS is a outer horizon and it is a inner horizon if $\pounds_{\xi} \Theta_k < 0$. It is easy to see this by setting up a locally Minkowski coordinate system at an event on the curve $\Theta_k=0$. The lie derivative along $\xi$ yields an invariant sign if the tangent to the curve $\Theta_k=c$ is null or timelike. Finding $\pounds_{\xi} \Theta_{k}$ at $\Theta_{k} = 0$ we get
    \begin{equation*}
    \pounds_{\xi} \Theta_{k} \bigg|_{\Theta_{k} = 0} = \frac{-n}{a\dot{a}^2}
    \end{equation*}
This could be further reduced by using the evolution of $a(t)$ from (\ref{homEE1}) for $k=0$ and using (\ref{homdensity}) with $\omega_{critical}$ we get
    \begin{equation} \label{reddegevol}
    a^2 \dot{a}^2 = \frac{\kappa \rho_{o}}{n(n+1)}
    \end{equation}
where $\rho_{o}$ is the initial mass density (which is always taken to be a positive constant). So finally we have
    \begin{equation}
    \pounds_{\xi} \Theta_{k} \bigg|_{\Theta_{k} = 0} = \frac{-\kappa \rho_{0}}{n+1} < 0
    \end{equation}
This implies that the future degenerate horizon (null MTS) is an inner horizon.
\\

The evolution of these degenerate horizons is obtained from the expression (\ref{reddegevol}) which gives (with the condition that $a(t=0)=1)$
    \begin{equation}
    a(t) = \pm \bigg( \frac{4 \kappa \rho_{o}}{n(n+1)} \bigg)^{1/4} (t+t_0)^{1/2} .
    \end{equation}
where $t_0$ is integration constant which is fixed by the initial condition. From this we can see that the physical radius $R = a(t) r$ of these degenerate horizons is always monotonically increasing or decreasing, the rate of which is dependent on the dimension of the spacetime $(n)$ and the initial mass density $(\rho_{o})$.

\section{Conclusions}

The causal nature of MTT and MATT is governed by the product of lie derivatives of $\Theta$'s with respect to both the null directions and also by ratio of these lie derivatives. These lie derivatives of $\Theta$'s are obtained just from the structure of the Einstein equations into a closed form expression in terms of the number of dimensions of spacetime $D=n+2$, cosmological constant $\Lambda$, principle pressure $p$ and density $\rho$ and physical radius $R$ which is a geometric quantity defined by the killing vectors of the spherical symmetry.
\\

The causal nature of MTT and MATT for the homogeneous case gave us causal phase portraits in the space of physical parameters of the model $(n,\Lambda,\omega,R)$. From these we were able to see causal transitions in ads, ds and flat spacetime. We obtain an interesting geometric relation where the spacetime is partitioned into distinct regions by an $R_{critical}$ based on causal behavior of evolving trapping horizons. We also saw that the causal transitions were also equivalent to inner and outer horizon transitions. 
\\

We found a few cases of degenerate horizons which are null everywhere and are also evolving. The degenerate horizon classification as inner or outer horizon described by Hayward was shown to be incomplete because of the vanishing of lie derivatives for all orders. Hence we made a prescription of taking the lie derivative of $\Theta$ along a tie-breaker vector field to carry out this classification.
\\

The analysis brings out few questions that we look to answer in our future works. When there is a (non-equilibrium) thermodynamic interpretation for evolving horizons the causal phases and there transitions may be useful. When one considers the cases where the MTT or MATT is null everywhere but also evolving then they form a third category of evolving horizons in addition to the dynamical horizons and timelike tubes, hence an area balance law for such degenerate horizons can be formulated.

%%%%%
\bibliographystyle{unsrt}
\bibliography{references}
%%%%%

\end{document}